\documentclass[12pt]{article}
\usepackage{amsmath}
\usepackage{amsthm}
\usepackage{amscd}
\usepackage{epsfig}
\usepackage{textcomp}
\usepackage{fullpage}
\usepackage{natbib}
\usepackage{amsfonts}
\usepackage[usenames,dvipsnames]{color}

\newcommand{\bm}[1]{\mbox{\boldmath $#1$}}
\newcommand{\mb}[1]{\mathbf{#1}}
\renewcommand{\Re}[0]{\mathbb{R}}

\newcommand{\NA}[0]{\mbox{\tt NA}}
\newcommand{\ith}[1]{$#1^{\mbox{\tiny th}}$}
\DeclareMathOperator*{\argmin}{argmin}

\newcommand{\gr}{\color{green}}
\newcommand{\rd}{\color{red}}

\begin{document}

\title{
  Shrinkage regression for multivariate inference with missing
  data, and an application to portfolio balancing} 
\author{
  Robert B. Gramacy\\
  Statistical Laboratory\\
  University of Cambridge\\
  bobby@statslab.cam.ac.uk \and 
  Ester Pantaleo\\
  Dipartimento di Fisica\\
  Universit\`{a} di Bari, Italy\\
  ester.pantaleo@ba.infn.it}

\maketitle

\begin{abstract}
  Portfolio balancing requires estimates of covariance between asset
  returns.  Returns data have histories which greatly vary in length,
  since assets begin public trading at different times.  This can lead
  to a huge amount of missing data---too much for the conventional
  imputation-based approach.  Fortunately, a well-known factorization
  of the MVN likelihood under the prevailing historical missingness
  pattern leads to a simple algorithm of OLS regressions that is much
  more reliable. When there are more assets than returns, however, OLS
  becomes unstable.  \cite{gra:lee:silva:2008} showed how classical
  shrinkage regression may be used instead, thus extending the state
  of the art to much bigger asset collections, with further accuracy
  and interpretation advantages. In this paper, we detail a fully
  Bayesian hierarchical formulation that extends the framework further
  by allowing for heavy-tailed errors, relaxing the historical
  missingness assumption, and accounting for estimation risk. We
  illustrate how this approach compares favorably to the classical one
  using synthetic data and an investment exercise with real returns.
  An accompanying {\sf R} package is on CRAN.

\end{abstract}

\section{Introduction}
\label{sec:intro}

\begin{figure}[ht!]
\centering
\includegraphics[scale=0.4,trim=0 10 35 0, clip=true]{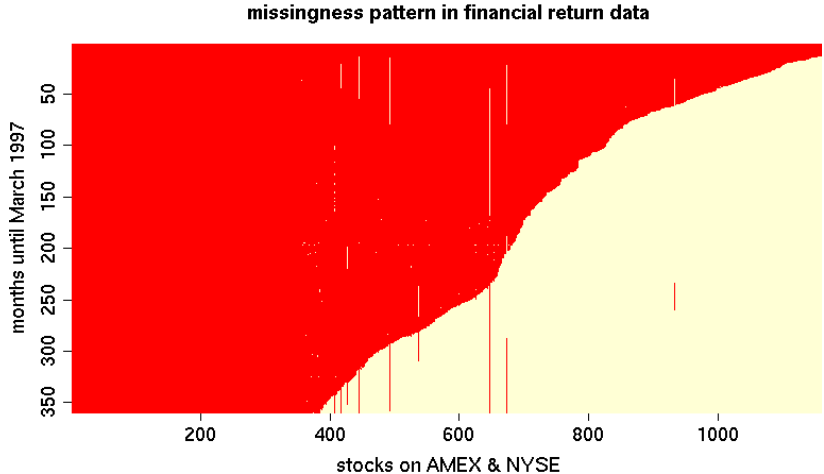} 
\hspace{0.2cm}
\caption{Missingness pattern in stock returns on the NYSE \& AMEX.  
The assets (columns) are put into increasing order of the number of
missing entries. }
\label{f:monoret}
\end{figure}

Mean--variance portfolio allocation \citep[e.g.,][]{markowitz:1959}
requires the accurate and tractable estimation of the mean return, and
the covariance between the returns, of a large number of assets.
Assets become publicly tradeable at different times, so their return
histories can greatly vary in length.  Aside from a few ``gaps'', the
histories of assets which are publicly tradeable at purchase time will
exhibit a {\em monotone missingness pattern}.  For example, Figure
\ref{f:monoret} shows the monthly return availability for 1,200-odd
stocks on NYSE \& AMEX over 29 years, some with as little as
1 years worth of data.
The assets along the columns have been sorted by the number of missing
entries.  Besides some ``gaps'' the boundary between the dark and
light regions is monotone.

Generally speaking, inference in the presence of missing data is
notoriously difficult, usually requiring hill-climbing iterative
techniques like {\em expectation maximization}
\citep{little:rubin:2002,schafer:1997}) which rapidly lose stability
as the level of missingness increases. The Bayesian alternative of
{\em data augmentation} is similarly unsatisfactory.  Software
packages implementing such algorithms come with prominently displayed
warnings of failure when the missingness level is above 15\%
\citep[see][]{gra:lee:silva:2008}.

The nice thing about a monotone missingness pattern is that the
likelihood has a convenient factorization which makes inference
tractable without imputation.  Under a multivariate normal (MVN)
assumption, a simple algorithm \citep{andersen:1957,stambaugh:1997} of
ordinary least squares (OLS) regressions, one for each asset, yields a
maximum likelihood estimator (MLE). Unfortunately, there must be fewer
stocks than the length of the shortest return history, so that the
design matrices of the OLS regressions are of full rank.  In
particular, you cannot have more stocks than historical returns.
\cite{gra:lee:silva:2008} showed that by replacing the OLS with
``parsimonious regressions'', e.g., principal components (PCR), ridge,
lasso, etc., the above algorithm can be applied when there are more
assets than historical returns. This extended the reach of Stambaugh's
(1997) methods from dozens to thousands of assets, accommodating an
essentially arbitrary level of historical missingness.

In this paper we shall further extend the above parsimonious
methodology in several directions by taking a fully Bayesian approach.
Section \ref{sec:monotone} recalls the monotone decomposition and
(MLE/Bayesian) inference algorithm.  Section \ref{sec:bshrink} reviews
approaches to Bayesian shrinkage regression that are particularly
convenient in this context, and which allow model averaging and
heavy-tailed errors as minor embellishments.  Section
\ref{sec:bmonomvn} details how the benefits of the Bayesian shrinkage
posteriors filter through to inference about the mean vector and
covariance matrix.  It features extensions for data augmentation to
deal with ``gaps'', and allows estimation risk to influence the
balanced portfolios---both of which were unavailable previously.  In
Section \ref{sec:app} we apply our methods in a Monte Carlo investment
exercise and show how they compare favorably to the classical
alternatives on real financial returns data.  The paper concludes with
a brief discussion in Section \ref{sec:discuss}.

The methods that are core to this paper are implemented in a fully
documented {\sf R} \citep{rproject} package called {\tt monomvn}
\citep{monomvn}, which is available for download on the Comprehensive
{\sf R} Archive Network (CRAN).

\section{Multivariate normal monotone missing data}
\label{sec:monotone}

We assume that the missingness mechanism is {\em missing completely at
  random} (MCAR).  In the case of historical asset returns this may be
a tenuous assumption, but it is convenient and common
\citep[e.g.,][]{stambaugh:1997}.  We work with a $n \times m$ data
matrix $\mb{Y}$ that collects the historical returns of the assets.
Denote $y_{i,j} = \NA$ if the \ith{i} sample (historical return) of
the \ith{j} covariate (asset) is missing; otherwise $y_{i,j} \in \Re$.
Formally speaking, the missingness pattern in $\mb{Y}$ is said to be
{\em monotone} [e.g., \citep[][Section 6.5.1]{schafer:1997} or
\citep[][Section 7.4]{little:rubin:2002}] if its columns can be
re-arranged so that $y_{i,j} \ne \NA$ whenever $y_{i,j+1} \ne \NA$.

We assume throughout that they are indeed arranged in this way, so
that when we define $n_j = \sum_{i=1}^n \mathbb{I}_{\{y_{i,j} \ne
  \,\mbox{\tt \footnotesize NA}\}}$ as the number of observed entries
in column $j$, for $j=1,\dots, m$, we have that $n\equiv n_1$ and $n_j
\geq n_{j+1}$.  Furthermore, the rows may be arranged according to the
same property ($y_{i,j} \ne \NA$ whenever $y_{i+1,j} \ne \NA$) without
loss of generality, so that when we define $\mb{y}_j \equiv
y_{1:n_j,j}$ we are collecting the entirety of the observed entries in
the \ith{j} column.
\begin{figure}[ht!]
\centering
\input{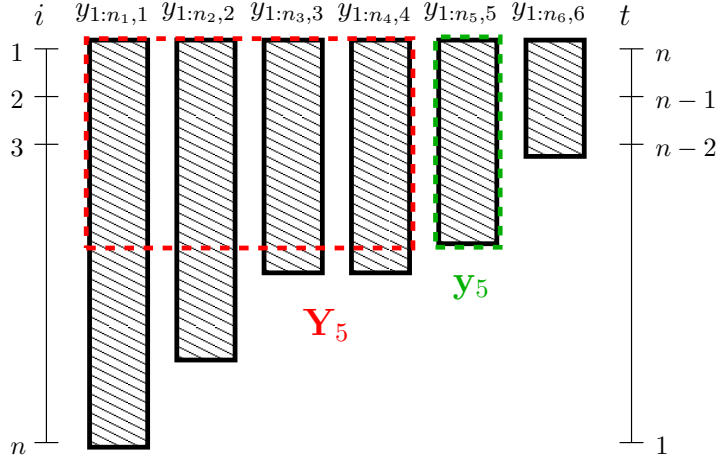}
\caption{ Diagram of a monotone missingness pattern with $m=6$
  covariates, and $n$ completely observed samples in
  $\mb{y}_1=y_{:,1}$.  The design matrix $\mb{Y}_5$ (without an
  intercept term) and the response vector $\mb{y}_5$ for the fifth
  regression involved in maximizing the likelihood of MVN data under a
  monotone missingness pattern is also shown.  Time $(t)$ runs counter
  to the index $i$ so that the most recent historical return is at
  time $t=n$, indexed by $i=1$.}
\label{f:monoreg}
\end{figure}
The pattern that results is illustrated pictorially in Figure
\ref{f:monoreg}.  For now we will assume that there are no ``gaps''.
It is also helpful to define $\mb{Y}_j \equiv
\mb{Y}_{0:(j-1)}^{(n_j)}$ as the $n_j \times j$ design matrix
\[
\mb{Y}_j \equiv \mb{Y}_{0:(j-1)}^{(n_j)} = \begin{pmatrix}
  1 & y_{1,1} & \cdots & y_{1,(j-1)} \\
  1 & y_{2,1} & \cdots & y_{2,(j-1)} \\
  \vdots & \vdots & \ddots & \vdots \\
  1 & y_{n_j,1} & \cdots & y_{n_j, (j-1)}
\end{pmatrix}
\]
containing an intercept, and the first $n_j$ rows of the first
$j-1$ columns of $\mb{Y}$; see Figure \ref{f:monoreg}.

Under the monotone pattern the likelihood $f(\mb{Y}|\bm{\theta})$
emits a convenient factorization in terms of an auxiliary
parameterization $\bm{\phi} = \Phi(\bm{\theta})$.  If the rows of
$\mb{Y}$ are i.i.d.~MVN, this factorization leads to an iterative
algorithm for inferring the MLE $\hat{\bm{\theta}} =
(\hat{\bm{\mu}},\hat{\bm{\Sigma}})$.  These assumptions may not be
appropriate for financial returns
but they are common to keep inference tractable
\citep[e.g.,][]{stambaugh:1997,ckl:1999,jagma:2003}.


The algorithm, due originally to \cite{andersen:1957}, begins by
calculating $\hat{\mu}_1$ and $\hat{\Sigma}_{1,1} \equiv
\hat{\sigma}_1^2$ in the usual way: $\hat{\mu}_1 = n_1^{-1}
\sum_{i=1}^{n_1} y_{i,1}$ and $\hat{\sigma}_1^2 = n_1^{-1}
\sum_{i=1}^{n_1} (y_{i,1} - \hat{\mu}_1)^2$.  Then, for $j=2,\dots,m$
the MLEs of $\bm{\theta}_j = (\mu_j, \bm{\Sigma}_{1:j,j})$,
$j=2,\dots,m$, can then be obtained via a regression on the complete
data in columns $1,\dots,j-1$, i.e., using the model $\mb{y}_j =
\mb{Y}_j \bm{\beta}_j + \bm{\epsilon}_j$, where
$\{\epsilon_{i,j}\}_{i=1}^{n_j} \stackrel{\mathrm{iid}}{\sim}
\mathcal{N}(0,\sigma_j^2)$.  Here $\bm{\beta}_j^\top = (\beta_{0,j},
\beta_{1,j}, \dots, \beta_{(j-1),j})$, and $\sigma_j^2$ are the
auxiliary parameters $\bm{\phi}_j$.  When $\mathrm{rank}(\mb{Y}_j) =
j$, and particularly when $n_j > j$, MLEs $\hat{\bm{\phi}}_j$ may be
obtained in the usual way: $\hat{\bm{\beta}}_j = (\mb{Y}_j^\top
\mb{Y}_j)^{-1} \mb{Y}_j^\top \mb{y}_j$ and $\hat{\sigma}^2_j =
\frac{1}{n_j} ||\mb{y}_j - \mb{Y}_j \hat{\bm{\beta}}_j||^2 =
\frac{1}{n_j} \sum_{i=1}^{n_j} (y_{i,j} - (\mb{y}_i^\top)_{1:n_j}\,
\hat{\bm{\beta}}_j)^2$.
The components of $\hat{\bm{\theta}}_j$ given
$\hat{\bm{\theta}}_{1:(j-1)} = (\hat{\bm{\mu}}_{1:(j-1)}^\top,
\hat{\bm{\Sigma}}_{1:(j-1),1:(j-1)})$ and 
$\hat{\bm{\phi}}_j$ are then
\begin{align}
  \hat{\mu}_j &= \hat{\beta}_{0,j} + \hat{\bm{\beta}}_{1:(j-1),j}^\top
  \hat{\bm{\mu}}_{1:(j-1)}
&\hspace{-0.075cm} \mbox{and}&&
\hat{\bm{\Sigma}}_{1:j,j}
  &= \begin{pmatrix}
    \hat{\bm{\beta}}_{1:(j-1),j}^\top \hat{\bm{\Sigma}}_{1:(j-1),1:(j-1)} \\
    \hat{\sigma}^2_j + \hat{\bm{\beta}}_{1:(j-1),j}^\top
    \hat{\bm{\Sigma}}_{1:(j-1),1:(j-1)} \hat{\bm{\beta}}_{1:(j-1),j},
\label{eq:addy}
\end{pmatrix}.
\end{align}
The $\hat{\bm{\Sigma}}$ thereby obtained will be positive-definite,
as long as $n_j > j$ for all $j=1,\dots,m$ so that
$\mb{Y}_j$ is of full rank, and $\mb{Y}_j^\top \mb{Y}_j$ invertible.

\subsection{Bayesian inference}
\label{sec:bayes}

The Bayesian approach follows naturally from priors on the auxiliary
parameters $\bm{\beta}_j$ and $\sigma_j^2$, for $j=1,\dots,m$.
Samples from the implied posterior of $\mu_j$ and
$\bm{\Sigma}_{1:j,j}$ are then obtained via $\Phi^{-1}$ in
Eq.~(\ref{eq:addy}).  It may be more desirable to choose priors
directly in the natural parameter space $\bm{\theta} = (\bm{\mu},
\bm{\Sigma})$, but it can be difficult to analytically derive the
implied priors for $\bm{\beta}_j$ and $\sigma_j^2$.  However, a
popular non-informative prior used for MVN data, $p(\bm{\mu},
\bm{\Sigma}) \propto |\bm{\Sigma}|^{-\left(\frac{m+1}{2} \right)}$,
can be shown \citep[][Section 6.5.3]{schafer:1997} to imply the
(independent) prior(s) $p(\bm{\beta}_j,\sigma_j^2) \propto
(\sigma_j^2)^{-\left(\frac{m+1}{2}-m+j \right)}$, giving the posterior
conditionals:
\begin{align*}
\bm{\beta}_j | \sigma^2_j, \bm{y}_j, \bm{Y}_j & 
\sim \mathcal{N}_{j}(\hat{\bm{\beta}}_j, \sigma^2_j (\bm{Y}_j^\top \bm{Y}_j)^{-1}) \\
\sigma^2_j | \mb{y}_j, \mb{Y}_j 
&\sim \mathrm{IG}((n_j-m+j-1)/2, 
(||\mb{y}_j - \mb{Y}_j \hat{\bm{\beta}}_j||^2)/2).
\end{align*} 
\cite{stambaugh:1997} showed that, under this non-informative prior,
it is possible to derive the moments of the Bayesian posterior {\em
  predictive} distribution in terms of the MLEs
$(\hat{\bm{\mu}},\hat{\bm{\Sigma}})$ in closed form.  When these are
used in the mean--variance framework to construct portfolios, they are
said to take {\em estimation risk} \citep{klein:bawa:1976,brown:1979}
into account. However, these calculations similarly break down when
$n_j \leq j$.

Inverted Wishart priors for $\bm{\Sigma}$ are amenable to tractable
posterior inference under the MVN with monotone missingness
\citep{liu:1993}.  One example is a {\em ridge prior} \citep[][Section
5.2.3]{schafer:1997}, which is helpful when $m > n$ and is closely
related to {\em ridge regression} [see Section \ref{sec:bshrink}] as
used by \cite{gra:lee:silva:2008} in $\bm{\phi}$-space to good effect.
This motivates a more pragmatic approach to prior selection: a
deliberate search for appropriate shrinkage priors for the ``big $p$
small $n$'' regression problem in $\bm{\phi}$-space, where the low
rank $\mb{Y}_j$ problem is manifest.  Then, $\Phi^{-1}$ completes the
description implicitly in $\bm{\theta}$-space.

\section{Bayesian shrinkage regression}
\label{sec:bshrink}

Here we focus on appropriate regression models for the ``big $p$ small
$n$'' problem that employ shrinkage.  The customary formulation is
\begin{align}
\mb{y} &= \beta_0 \mb{1}_n + \mb{X}\bm{\beta} + \bm{\epsilon},
&& \mbox{where} &
\mb{\epsilon} &\sim \mathcal{N}_n(\mb{0}, \sigma^2 \mb{I}_n).
\label{eq:br}
\end{align}
One typically assumes a standardized $n \times p$ design matrix
$\mb{X}$ where the columns are individually adjusted to have zero-mean
and unit L2-norm.  This causes $\beta_0$ and $\bm{\beta}$ to
be independent {\em a posteriori} and recognizes that regularized
posterior summaries for $\bm{\beta}$ are not equivariant under a
re-scaling of $\mb{X}$.  Any such pre-processing must be undone before
evaluating $\bm{\theta} = \Phi^{-1}(\bm{\phi})$ at samples of
$\bm{\phi} = (\beta_0, \bm{\beta}, \sigma^2)$.


{\em Ridge regression} and the {\em lasso} \citep[e.g.,][Section
3.4.3]{hastie:tibsh:fried:2001} are classical approaches to shrinkage
regression that penalize large coefficients:
\begin{equation}
  \hat{\bm{\beta}}^{(q)} = \argmin_{\bm{\beta}}
  \left\{(\tilde{\mb{y}} - 
\mb{X}\bm{\beta})^\top(\tilde{\mb{y}} - \mb{X}\bm{\beta}) + 
    \lambda \sum_{j=1}^p |\beta_j|^q\right\}
\label{eq:qp}
\end{equation}
for some $\lambda \geq 0$, where the intercept is excluded from
penalization via $\tilde{\mb{y}} = \mb{y} - \bar{y}\mb{1}_n$.
Choosing $q=2$ yields {\em ridge regression} \citep{hoerl:1970} where
$\hat{\bm{\beta}}^{(2)} = (\mb{X}^\top \mb{X} + \lambda \mb{I})^{-1}
\mb{X}^\top \mb{y}$.  The {\em lasso} \citep{tibsh:1996} corresponds
to $q=1$. There is no closed form solution for
$\hat{\bm{\beta}}^{(1)}$, but the entire path of solutions for all
$\lambda$ can be obtained iteratively via the LARS algorithm
\citep{efron:2004}.  Both estimators may be interpreted as the
posterior mode under a particular prior.  For ridge regression the
prior is $\bm{\beta}^{(2)}|\sigma^2 \sim \mathcal{N}_p(\mb{0},
\sigma^2 \lambda \mb{I}_p)$; for the lasso it is i.i.d.~Laplace (i.e.,
double-exponential)
$
\pi(\bm{\beta}^{(1)}| \sigma^2) = \prod_{j=1}^p 
\frac{\lambda}{2\sqrt{\sigma^2}} e^{-\lambda |\beta^{(1)}_j|/\sqrt{\sigma^2}}.
$

Large values of the penalty parameter $\lambda$ cause the coefficients
of $\hat{\bm{\beta}}^{(q)}$ to be shrunk towards zero.  The lasso
estimator $\hat{\bm{\beta}}^{(1)}$ may have many coefficients shrunk
to exactly zero, which is convenient for variable selection.  Often,
$\lambda$ is chosen via cross validation (CV).  As a
$\bm{\phi}$-space regression for obtaining monotone MVN estimators.
\cite{gra:lee:silva:2008} chose $\lambda$ by applying the
``one-standard-error'' rule \citep[][Section
7.10]{hastie:tibsh:fried:2001} with CV.

\subsection{Hierarchical models for Bayesian shrinkage regression}
\label{sec:blasso}

For a fully Bayesian lasso we use the latent variable formulation of
\cite{park:casella:2008} and \cite{carlin:polson:1991} by representing
the Laplace as a scale mixture of normals:
\begin{align}
  \mb{y} | \beta_0, \mb{X}, \bm{\beta}, \sigma^2 & \label{eq:model}
  \sim \mathcal{N}_n(\beta_0 \mb{1}_n + \mb{X} \bm{\beta}, \sigma^2
  \mb{I}_n)
  & \mb{D}_\tau &= \mathrm{diag}(\tau_1^2, \dots, \tau_p^2) \\
  \bm{\beta} | \sigma^2, \tau_1^2,\dots, \tau_p^2 & \sim
  \mathcal{N}_p(\mb{0}, \sigma^2 \mb{D}_\tau), \;\; \beta_0 \propto 1
  &
  \sigma^2 &\sim \mathrm{IG}(a_\sigma/2, b_\sigma/2) \nonumber \\
  \tau_j^2|\lambda & \nonumber \stackrel{\mathrm{iid}}{\sim}
  \mathrm{Exp}(\lambda^2/2) & \lambda^2 &\sim G(a_\lambda, b_\lambda).
  \nonumber
\end{align}
IG and G are the rate- and scale-parameterized inverse-gamma and
gamma distributions, respectively.  The default prior $\pi(\sigma^2)
\propto \sigma^{-2}$ is obtained with $a_\sigma = b_\sigma = 0$, and
ridge regression is the special case where $\tau^2 \equiv \tau_1^2 =
\cdots = \tau_p^2$, using $\tau^2\sim\mathrm{IG}(a_\tau/2, b_\tau/2)$
(i.e., dropping $\lambda^2$), possibly with $a_\tau = b_\tau = 0$.
Fixing $\tau^2 = \infty$ yields the standard family of (improper)
priors for linear regression.

Choosing the prior parameterization for $\lambda^2$ can be difficult.
\cite{park:casella:2008} note that choosing $a_\lambda = b_\lambda =
0$ leads to an improper posterior, and suggest some automatic
alternatives.  Another option is to further expand the hierarchy using
a so-called normal-gamma (NG) prior for $\beta$
\citep[e.g.,][]{griffin:brown:2010} by specifying
\begin{align}
  \lambda^2 |\gamma &\sim G(a_\lambda, b_\lambda/\gamma),
  \;\;\;\;\; \mbox{where}\;\;\;\;\;
  \gamma \sim \mathrm{Exp}(1) \label{eq:ng}\\
 \mbox{and} \;\;\;\;\; \tau_j^2|\lambda^2, \gamma
 &\stackrel{\mathrm{iid}}{\sim}  G(\gamma, \lambda^2/2). \nonumber
\end{align}
\cite{griffin:brown:2010} suggest $a_\lambda = 2$, and $b_\lambda =
M/2$, where $M$ is chosen via empirical Bayes considerations.  Observe
that fixing $\gamma = 1$ encodes the specific Laplace prior case.  So
the NG prior is more adaptive than the lasso.  This may come in handy
when $p \gg n$, i.e., when our prior plays a more important role, or
when the posterior drives many $\beta_j$'s to zero.

The availability of full conditionals for all of the parameters makes
for efficient Gibbs sampling (GS).  For the baseline Bayesian lasso
model \citep{park:casella:2008} these are:
\begin{align}
\beta_0 | \sigma^2, \mb{y} &\sim \mathcal{N}(\bar{y}, \sigma^2/n) \nonumber \\
\bm{\beta}|\sigma^2, \{\tau_j^2\}_{j=1}^p, \mb{y} 
&\sim \mathcal{N}_p(\tilde{\bm{\beta}}, \sigma^2 \mb{A}^{-1}),
& \mb{A} &= \mb{X}^\top \mb{X} + \mb{D}_\tau^{-1}, 
\;\; \tilde{\bm{\beta}} = \mb{A}^{-1} \mb{X}^\top \tilde{\mb{y}} \nonumber \\
\sigma^2| \bm{\beta}, \{\tau_j^2\}_{j=1}^p, \mb{y} &\sim 
\mathrm{IG}((a_\sigma + n-1+p)/2, 
   (b_\sigma + \psi_\beta)/2),
& \psi_\beta &= ||\tilde{\mb{y}}-\mb{X}\bm{\beta}||^2
+ \bm{\beta}^\top \mb{D}_\tau^{-1} \bm{\beta} \nonumber \\
\tau_j^{-2}| \beta_j, \sigma^2, \lambda &\stackrel{\mathrm{iid}}{\sim} 
\mbox{Inv-Gauss}(\sqrt{\lambda^2\sigma^2/\beta_j^2}, \lambda^2)
 \label{eq:postcond}  \\
\lambda^2 | \tau_1^2, \dots, \tau_p^2 &\sim 
\mbox{G}(a_\lambda + p\gamma, b_\lambda/\gamma + \textstyle
\sum_{j=1}^p \tau_j^2/2). && [\mbox{assuming } \gamma = 1]
\nonumber
\end{align}
Using a marginal posterior conditional for $\sigma^2$ instead can 
help reduce autocorrelation in the Markov chain.  
Integrating over the posterior conditional for $\bm{\beta}$ gives $
\sigma^2 | \tau_1^2, \dots, \tau_p^2, \mb{y}\sim\mathrm{IG}((a_\sigma
+ n - 1)/2, (b_\sigma + \psi_{\tilde{\beta}})/2), $ where
$\psi_{\tilde{\beta}} = ||\tilde{\mb{y}}-\mb{X}\tilde{\bm{\beta}}||^2
+ \tilde{\bm{\beta}}^\top \mb{D}_\tau^{-1} \tilde{\bm{\beta}} =
\tilde{\mb{y}}^\top \tilde{\mb{y}} - \tilde{\bm{\beta}}^\top \mb{A}
\tilde{\bm{\beta}}$.  

Under the ridge regression model the posterior conditionals are the
same (\ref{eq:postcond}) except that we ignore $\lambda^2$ and take
$\tau^2 \sim \mathrm{IG}((a_\tau + p)/2,(b_\tau +
\sigma^{-2}\bm{\beta}^\top \bm{\beta})/2)$.  Upon fixing $\tau^2 =
\infty$ we must use $\mb{A} = \mb{X}^\top \mb{X}$, subtract $p/2$ to
the rate parameter to the IG conditional(s) for $\sigma^2$, and ensure
a proper posterior with $a_\sigma > p-n-1$.  This may pose a
non-trivial restriction on the prior when $p \geq n$.  Under the NG
prior $\gamma$ may vary, leading to the conditionals
\begin{align}
\tau_j^{2}| \beta_j, \sigma^2, \lambda^2, \gamma &\stackrel{\mathrm{iid}}{\sim} 
\mathrm{GIG}(\gamma-1/2,\beta_j^2/\sigma^2, \lambda^2) \\
\gamma | \{\tau_j^2\}_{j=1}^p, \lambda^2 \nonumber
&\propto \left( \frac{\lambda^2}{2}\right)^{p\gamma} 
\frac{\pi(\gamma)}{(\Gamma(\gamma))^p}\left( \prod_{i=1}^p{\tau_j^2} \right)^{\gamma},
\end{align}
where $\mathrm{GIG}(\lambda, \chi, \psi)$ is the generalized inverse
Gaussian distribution.  \cite{griffin:brown:2010} suggest a random
walk Metropolis update for the $\gamma$ using proposals $\gamma' =
\exp\{\sigma_\gamma z\}$, for $z\sim \mathcal{N}(0,1)$.  These proposals are
accepted with probability
\begin{equation}
\min\left\{1,\frac{\pi(\gamma')}{\pi(\gamma)} 
\left( \frac{\Gamma(\gamma)}{\Gamma(\gamma')} \right)^p 
\left(\left(\frac{2}{\lambda^2}\right)^{-p} 
\prod_{i=1}^p \tau_j^2 \right)^{\gamma'-\gamma}\right\},
\end{equation}
where $\pi(\gamma)=\gamma^{-2}
\exp(-\gamma-\frac{M}{2\gamma}\lambda^2)$ and $\sigma_\gamma$ is
chosen to give an acceptance rate of 20-30\%.  

An alternative hierarchical modeling framework for the Bayesian lasso
is provided by \cite{hans:2008}.  While it does not require $p$ latent
$\tau^2_j$ variables, the resulting GS procedure is
not fully blocked, and rejection sampling is required for $\sigma^2$.
``Orthogonalizing'' the sampler helps mitigate slow mixing of
the un-blocked conditionals.  However, we prefer the simpler approach of
\cite{park:casella:2008} as it is more readily adaptable to the $p \gg
n$ case, to model selection, and our extensions to the heavy-tailed
errors.

Whereas the classical lasso has the property that the estimate
$\hat{\bm{\beta}}^{(1)}$ may have components which are zero---in fact,
it would never have more than $\min\{p,n-1\}$ nonzero components---
samples of $\bm{\beta}$ from the posterior would never have zeros.  So
the Bayesian lasso is less useful for variable selection.  We also
note that when $p \geq n$---and without the ability to explicitly
restrict $\bm{\beta}$ to having at most $\min\{p,n-1\}$ nonzero
components---a proper prior must be used for $\sigma^2$ or the
posterior will be improper.  An empirical Bayes remedy that works well
in this case is to take a small $a_\sigma$, say $a_\sigma = 3/2$, and
then set $b_\sigma$ so that the $(1-\alpha)$ part of the
IG($a_\sigma,b_\sigma)$ distribution lies at the point
$\tilde{\mb{y}}^\top \tilde{\mb{y}}$ (i.e., the MLE under the
intercept model) via the incomplete gamma inverse function.
Another remedy is Bayesian model averaging.

\subsection{Bayesian model selection and averaging}
\label{sec:bma}

Although the MAP lasso fit may indeed set some of the coordinates of
$\hat{\bm{\beta}}^{(1)}$ to zero, this is more of a side effect of the
solution space of the quadratic program (\ref{eq:qp}) than the result
of a deliberate prior modeling choice \citep{hans:2008}.  Bayesians
rarely base inference on the MAP; it is more natural to select
variables by inspecting the posterior model probabilities.

There are several standard ways of performing Bayesian variable
selection in regression models that are amenable to GS.  They
essentially fall into two camps. Loosely, the first camp
\citep[e.g.,][]{geweke:1996,georg:mccu:1993} uses a product-space
wherein the prior for each $\beta_j$ is augmented to include a
point-mass at zero.  Inference proceeds by GS on each of the
conditionals $\beta_j| \bm{\beta}_{-j}, \mb{y}, \dots$, $j=1,\dots,p$,
which may flop between zero and nonzero values.  \cite{hans:2008}
augmented this product space approach to variable selection under the
Laplace prior by further conditioning on $\lambda$.

The second camp \citep[e.g.,][]{troughton:godsill:1997} is
transdimensional in that the $\bm{\beta}$-vector may vary in length
while model space is traversed via Reversible Jump (RJ) MCMC
\citep{gree:1995}.  We prefer this approach for our monotone inference
application in the \cite{park:casella:2008} setup.  When $p \gg n$ it
is implementationally more compact, only requiring memory for the
(nonzero) $\bm{\beta}$-components.  This represents a big savings when
simultaneously storing $m$ regression model parameter sets under a
prior that restricts the model to have at most $\min\{j,n_j-1\}$
(nonzero) coefficients.  Also, we like the fully blocked samples for
the nonzero components of $\bm{\beta}$ for the within-model moves.

Suppose that the transdimensional Markov chain is currently visiting
some model with $k$ nonzero regression coefficients $\bm{\beta}_{k} =
(\beta_1,\dots,\beta_k)$ using design matrix $\mb{X}_k$.  The columns
of $\mb{X}_k$ should come from a two-way partition (of $k$ and $p-k$
elements) of the $p$ columns of $\mb{X}$, but they need not coincide
with the first $k$ of the $p$ columns.  Now consider proposing to add
a column to $\mb{X}_k$, a so-called ``birth'' move.  Choose one of
the $p-k$ columns of $\mb{X}$ not present in $\mb{X}_k$ for addition,
thus creating $\mb{X}_{k+1}$.  By considering the ratio of the
marginal posterior distributions (integrating out $\bm{\beta}_k$ and
$\bm{\beta}_{k+1}$) conditional on $\sigma^2, \tau_1^2,\dots,\tau_k^2$
and a new proposed $\tau_{k+1}^2$ (which we take from the prior), it
can be shown that the transdimensional move may be accepted with
probability $\min\{1, A_{k\rightarrow k+1}\}$, where
\begin{equation}
  A_{k\rightarrow k+1} = \frac{(\tau_{k+1}^{-2}|\mb{A}_{k+1}^{-1}|)^{\frac{1}{2}}
\exp\left\{\frac{1}{2\sigma^2}\tilde{\bm{\beta}}_{k+1}^\top 
   \mb{A}_{k+1}\tilde{\bm{\beta}}_{k+1}\right\}}{
|\mb{A}_k^{-1}|^{\frac{1}{2}}
\exp\left\{\frac{1}{2\sigma^2}\tilde{\bm{\beta}}_k^\top 
\mb{A}_k\tilde{\bm{\beta}}_k\right\}q(\tau_{k+1}^2)} \times 
\frac{\pi(k+1)q(k+1 \rightarrow k)}{\pi(k) q(k \rightarrow k+1)},
\label{eq:bma}
\end{equation}
and $\mb{A}_k = \mb{X}_k^\top \mb{X}_k + \mb{D}_{\tau_k}^{-1}$,
$\tilde{\bm{\beta}}_k = \mb{A}_k^{-1} \mb{X}_k^\top \tilde{\mb{y}}$,
with $\mb{D}_{\tau_k} = \mathrm{diag}(\tau_1^2,\dots, \tau_k^2)$.  The
reverse ``death'' move, of proposing to remove one of the columns of
$\mb{X}_{k}$, may be accepted with probability $\min\{1,
A_{k-1\rightarrow k}^{-1}\}$.  Under the ridge prior, $\tau_{k+1}^2 =
\tau^2$ can be dropped from the expression unless $k = 0$; for
standard regression it may be ignored so long as a proper prior is
used for $\bm{\beta}_k$.

A uniform prior over all models with $k$ nonzero components is
typical.  Often, $\pi(k) \propto 1, \forall k\in\{1,\dots,p^*\}$.
However, we prefer to take $k \sim \mathrm{Bin}(p^*,\pi)$, with $\pi
\in (0,1)$ where $\pi$ controls the ``sparsity'', and $p^*$ denotes
$p$ or $\min\{p,n-1\}$ for compactness.  Prior information on $\pi$
may be interjected either by fixing a particular value, or by taking a
hierarchical approach with $\pi \sim \mathrm{Beta}(g,h)$
\citep[e.g.,][]{georg:mccu:1993}.  \cite{hans:2008} used $g=h=1$ for
with a Laplace prior in the product space.  The posterior conditional
for GS is $\pi|k \sim \mathrm{Beta}(g+k, g+ p^*-k)$.  In our
transdimensional approach, we choose a uniform proposal for the valid
jumps. For a ``birth'' we take $q(0 \rightarrow 1) = 1/p$, and $q(k
\rightarrow k+1) = 1/2(p-k)$ for $k=1,\dots,p^*-1$.  Conversely for a
``death'' we take $q(p^* \rightarrow p^*-1) = 1/p^*$ and $q(k \rightarrow
k-1) = 1/2k$ for $k=p^*-1,\dots,1$.  Otherwise $q(k \rightarrow k') =
0$.

Movement throughout the $2^p$ sized space will be slow for large $p$,
so a certain amount of thinning of the RJ-MCMC chain is appropriate.
Collecting a sample from the posterior after $p$ transdimensional
moves approximates the model-level mixing (and computation burden) of
the product-space approach.  Throughout the RJ-MCMC the length of
$\bm{\beta}$ varies, and the components shift to represent the
partition of $\mb{X}$ stored in the columns of $\mb{X}_k$.  Therefore,
post-processing is necessary if samples of $\bm{\beta}$ are to be used
elsewhere, e.g., in $\bm{\theta}$-space via $\Phi^{-1}$
(\ref{eq:addy}), where a full $p$-vector having zero and nonzero
entries in the correct positions is needed.  This may be facilitated
by maintaining a $k$-vector of column indicators.  The posterior
probability that variable $j$, $j=1,\dots, p$, is relevant for
predicting $\mb{y}$ is then proportional to $\sum_{t=0}^T
\mathbb{I}_{\{\beta_j^{(t)} \ne 0\}}$, where $T$ is the number of
samples saved from the Markov chain.

\subsection{Student-$t$ errors via scale-mixtures}
\label{sec:scalemix}

The MVN assumption is not always appropriate.  We may wish to consider
the possibility that errors in $\mb{y}$ have a Student-$t$
distribution with an unknown degrees of freedom $\nu$:
\begin{align}
  \mb{y} &= \beta_0 \mb{1}_n + \mb{X} \bm{\beta} + \bm{\epsilon}, &
  \{\epsilon_i\}_{i=1}^n & \stackrel{\mathrm{iid}}{\sim}
  \mathrm{St}(0,\sigma^2;\nu).
 \label{eq:treg}
\end{align}
Following \cite{carlin:polson:stoffer:1992} and \cite{geweke:1993} we
shall represent the Student-$t$ distribution as a scale mixture of
normals with an IG$(\nu/2,\nu/2)$ mixing density.

We must redefine $\mb{X}=(\mb{1}_n,\mb{X})$ as a $n \times (p+1)$
matrix,
$\bm{\beta}=(\beta_0,\bm{\beta}^\top)^\top=\{\beta_j\}_{j=0}^p$ so
that the model in Eq.~(\ref{eq:treg}) becomes $\mb{y} = \mb{X}
\bm{\beta} + \bm{\epsilon}$ since the posterior intercept
$\beta_0$ is no longer independent of the other components of
$\bm{\beta}$ in the presence of heavy-tailed errors.  The setup is
otherwise unchanged from Section \ref{sec:blasso}.  Upon assuming an
exponential prior for the degrees of freedom parameter, $\nu$, the
modifications to the hierarchical model in Eq.~(\ref{eq:model}) are:
\begin{align}
  \mb{y} | \mb{X}, \bm{\beta}, \sigma^2, \{\omega_i^2\}_{i=1}^n & \sim
  \mathcal{N}_n(\mb{X} \bm{\beta}, \sigma^2 \mb{D}_\omega) &
  \mb{D}_\omega &= \mathrm{diag}(\omega_1^2, \dots, \omega_n^2)  \label{eq:tmodel} \\
  \bm{\beta}| \sigma^2,\{\tau_j^2\}_{j=1}^p & \sim \mathcal{N}_{p+1}(\mb{0},
  \sigma^2 \mb{D}_\tau)
  &  \mb{D}_\tau &= \mathrm{diag}(\infty,\tau_1^2, \dots, \tau_p^2) \nonumber \\
  \omega_i^2 | \nu & \nonumber \stackrel{\mathrm{iid}}{\sim}
  \mathrm{IG}(\nu/2, \nu/2) & \nu|\theta & \sim
  \mathrm{Exp}(\theta). \nonumber
\end{align}
Note that $\mb{D}_\tau$ is a $p+1$ diagonal matrix, and that the first
component insures that $\beta_0$ is given a flat prior as before.
After redefining $\mb{A} = \mb{X}^\top \mb{D}_\omega ^{-1} \mb{X} +
\mb{D}_\tau^{-1}, \tilde{\bm{\beta}} = \mb{A}^{-1} \mb{X}^\top
\mb{D}_\omega^{-1} \mb{y}$ and $\psi_{\beta} =
(\mb{y}-\mb{X}\bm{\beta})^\top
\mb{D}_\omega^{-1}(\mb{y}-\mb{X}\bm{\beta}) + \bm{\beta}^\top
\mb{D}_\tau^{-1} \bm{\beta}$, the modified full posterior
conditionals follow:
\begin{align}
\bm{\beta}|\sigma^2, \{\tau_j^2\}_{j=1}^p, \{\omega_i^2\}_{i=1}^n, \mb{y} 
&\sim \mathcal{N}_{p+1}(\tilde{\bm{\beta}}, \sigma^2 \mb{A}^{-1}) \label{eq:tpostcond} \\
\sigma^2 | \bm{\beta}, \{\tau_j^2\}_{j=1}^p, \{\omega_i^2\}_{i=1}^n, \mb{y} &
\sim \mathrm{IG} \left( \frac{a_\sigma + n + p}{2}, 
  \frac{b_\sigma + \psi_\beta}{2} \right) \nonumber   \\
\omega_i^2 | \bm{\beta}, \sigma^2, \nu, \mb{y} &
\stackrel{\mathrm{iid}}{\sim} \mathrm{IG} 
\left( \frac{\nu + 1}{2}, \frac{\nu + \sigma^{-2} ((\mb{y}-\mb{X} \bm{\beta})_i)^2}{2} 
\right) \nonumber \\
p(\nu|{\{\omega_i^2\}_{i=1}^n},\theta) &
\propto \left( \frac{\nu}{2} \right) ^{\frac{n\nu}{2}}
\left(\Gamma \left( \frac{\nu}{2} \right)\right) ^{-n} \exp({-\eta\nu}) \nonumber
\end{align}
where $\eta =
\frac{1}{2}\sum_{i=1}^{n}(\log(\omega_i^2)+\omega_i^{-2})+\theta$.\footnote{Note
  that there is a typo in the conditional for $\nu$ provided by
  \cite{geweke:1993}.}  

The conditional posterior of $\nu$ does not correspond to a standard
distribution, however a convenient rejection sampling method (with low
rejection rate) is available \citep{geweke:1992} using an exponential
envelope.  The optimal scale parameter $\nu^*$ can be chosen to
minimize the unconditional rejection rate by finding the root of
$(n/2) [\log( \nu/2) +1-\Psi(\nu/2)]+\nu^{-1}-\eta$, where $\Psi$ is
the digamma function. Standard Newton-like methods work well.  A draw
from $\nu \sim \mathrm{Exp}(\nu^*)$ may then be retained
with probability\footnote{There is also a typo in the acceptance
  probability provided by \cite{geweke:1992}.}
\[
\min\left\{1, \left[\frac{\Gamma(\nu^*/2)}{ \Gamma (\nu/2)}\right]^n
\left[\frac{(\nu/2)^{\nu}}{(\nu^*/2)^{\nu^*}}\right]^{n/2}
 \exp[(\nu-\nu^*)((\nu^*)^{-1}-\eta)]\right\}.
\]

As before we may integrate out $\bm{\beta}$
obtaining $\sigma^2 | \{\tau_j^2\}_{j=1}^p, \{\omega_i^2\}_{i=1}^n, \mb{y}
\sim \mathrm{IG}((a_\sigma + n-1)/2, (b_\sigma +
\psi_{\tilde{\beta}})/2)$ by redefining $\psi_{\tilde{\beta}} =
(\mb{y}-\mb{X}\tilde{\bm{\beta}})^\top
\mb{D}_\omega^{-1}(\mb{y}-\mb{X}\tilde{\bm{\beta}}) +
\tilde{\bm{\beta}}^\top \mb{D}_\tau^{-1} \tilde{\bm{\beta}} =
\mb{y}^\top \mb{D}_\omega^{-1}\mb{y} - \tilde{\bm{\beta}}^\top \mb{A}
\tilde{\bm{\beta}}$.  Finally, the Bayesian model selection and
averaging method of Section \ref{sec:bma}, via Eq.~(\ref{eq:bma}), may
be used with $\mb{X}_k=(\mb{1}_n,\mb{X}_k)$,
$\bm{\beta}_k=(\beta_0,\bm{\beta}_k^\top)^\top$, $\tilde{\bm{\beta}}
_k= \mb{A}_k^{-1} \mb{X}_k^\top \mb{D}_{\omega}^{-1} \mb{y}$ and
$\mb{A}_k=\mb{X}_k^\top \mb{D}_{\omega} ^{-1} \mb{X}_k +
\mb{D}_{\tau_k} ^{-1}$ and $\mb{D}_{\tau_k} =
\mathrm{diag}(\infty,\tau_1^2, \dots, \tau_k^2)$.  The number of
latent variables now grows with the sample size, so automatic $O(n)$
thinning from the Markov Chain is sensible.

\subsection{Empirical results on detecting fat tails}
\label{sec:blassores}

\cite{hans:2008} and \cite{griffin:brown:2010} offer a plethora of
insights about the Bayesian lasso and NG with comparison to the
classical lasso.  There is no need to re-produce these results here.
Instead we offer a demonstration of the Student-$t$ extensions that
are unique to our setup and relevant in light of the recent criticism
of MVN in the financial press. Basically, we explore the extent to
which deviations from normality may be detected by testing the null
hypothesis (model $\mathcal{M}_{\mathrm{N}}$) of normal errors versus
the alternative (model $\mathcal{M}_{\mathrm{St}}$) that they follow a
Student-$t$ with $\nu$ unknown.  One way to do this is via a posterior
odds ratio (POR):
\[
\frac{p(\mathcal{M}_{\mathrm{N}}|\mb{y})}{p(\mathcal{M}_{\mathrm{St}}|\mb{y})}
=\frac{\pi(\mathcal{M}_{\mathrm{N}})}{\pi(\mathcal{M}_{\mathrm{St}})} \times
\frac{p(\mb{y}|\mathcal{M}_{\mathrm{N}})}{p(\mb{y}|\mathcal{M}_{\mathrm{St}})} 
\equiv
[\mbox{prior ratio}] \times [\mbox{Bayes factor}]
\]
where $\pi(\mathcal{M}_{*})$ is the prior on $\mathcal{M}_{*}$, and
$p(\mb{y}|\mathcal{M}_{*})$ is the marginal likelihood for
$\mathcal{M}_{*}$.  By taking equal priors we may concentrate on the
Bayes factor (BF).

Calculating PORs and BFs can be difficult in generality; for a review
of related methods see \cite{godsill:2001}.  However, we may exploit
that the Student-$t$ and normal models differ by just one parameter in
the likelihood, $\nu$.  \citet[][Section
2.5.1]{jacq:polson:rossi:2004} show that this BF may be calculated by
writing it as the expectation of the ratio of un-normalized posteriors
with respect to the posterior under the Student-$t$ model.  That is,
we may calculate
\begin{align*}
\mathbb{E}\left\{ \frac{p(\mb{y}| \bm{\psi}, \mathcal{M}_{\mathrm{N}})}{
 p(\mb{y}|\bm{\psi}, \nu, \mathcal{M}_{\mathrm{St}})} \right\}
&\approx \frac{1}{T}\sum_{t=1}^T 
\frac{p(\mb{y}| \bm{\psi}^{(t)}, \mathcal{M}_{\mathrm{N}})}{
 p(\mb{y}|\bm{\psi}^{(t)}, \nu^{(t)}, \mathcal{M}_{\mathrm{St}})}, 
&& \mbox{where} &
(\bm{\psi}^{(t)}, \nu^{(t)}) &\sim 
p(\bm{\psi}, \nu | \mb{y}, \mathcal{M}_{\mathrm{St}}),
\end{align*}
and where $\bm{\psi}$ collects the parameters shared by both models.

To shed light on the ``selectability'' of the Student-$t$ model
(\ref{eq:treg}), consider synthetic data where $\bm{\beta} = (2, -3,
0, 0.75, 0, 0, -0.9)^\top$, $\mu\equiv \beta_0 = 1$, the rows of the
$n\times 7$ design matrix $\mb{X}$ are uniformly distributed in
$[0,1]^7$, and $\epsilon_i \sim \mathrm{St}(0,\sigma^2=1;\nu)$, for
$i=1,\dots, n$.  We perform a Monte Carlo experiment where $n$ and
$\nu$ vary, with $n\in\{30,75,100,200,500,1000\}$ and
$\nu\in\{3,5,7,10,\infty\}$, and consider the frequency of times that
the BF indicated ``strong'' preference for the correct model in
repeated trials.  In each trial, GS (\ref{eq:tpostcond}) was used to
obtain 1200 samples from the posterior by thinning every $7n$ rounds,
with the first 200 discarded as burn-in.  For $n\leq 200$ we repeated
the experiment with random data 300 times; when $n=500$ we used 50
replications; and when $n=1000$ we used 20.

\begin{figure}[ht!]
\centering
\includegraphics[scale=0.75,trim=0 25 0 35]{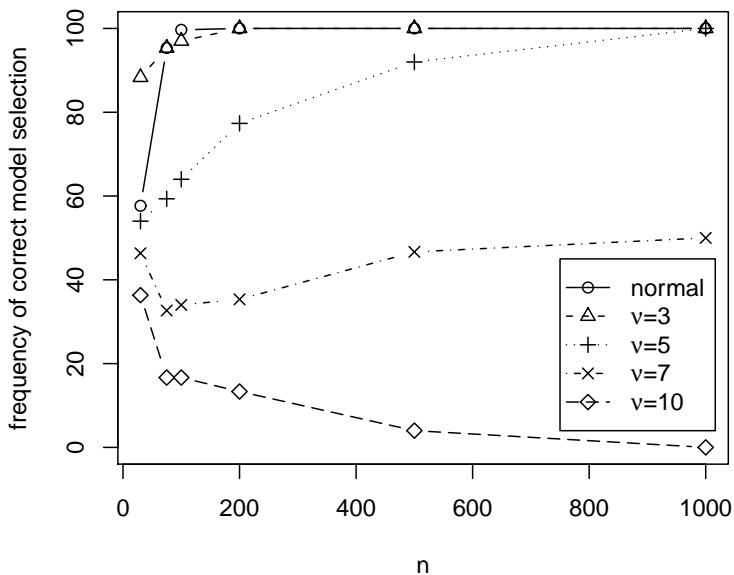}
\caption{Frequency of correct model determinations as a function of
  the sample size, $n$, and the degrees of freedom parameter, $\nu$,
  where ``normal'' is interpreted as $\nu = \infty$.}
\label{fig:bfnun}
\end{figure}
Figure \ref{fig:bfnun} shows the relationships between $n$, $\nu$ and
the frequency of correct model determinations (higher frequencies are
better).  In the case of normal errors, and Student-$t$ errors with
$\nu = 3$, the correct model can be determined with high accuracy when
$n\geq 200$.  When $\nu = 5$ a sample size of $n = 1000$ is needed;
when $\nu = 7,10$ we need $n\gg 1000$.  Clearly for $10 \leq \nu <
\infty$ the situation is hopeless unless $n$ is very large.  These
results have implications in the context of our motivating financial
returns data in Section \ref{sec:app}, indicating that long return
histories may be required to benefit from relaxing the MVN assumption.

\section{Bayesian inference under monotone missingness}
\label{sec:bmonomvn}

Here we collect ideas from the previous sections in order to sample
from the joint posterior distribution of $\bm{\theta} = (\bm{\mu},
\bm{\Sigma})$. Let $\bm{\phi}_j = (\beta_{0,j}, \bm{\beta}_j,
\sigma_j^2) \sim \mathcal{BR}_j \equiv \mathcal{BR}(\mb{Y}_j,
\mb{y}_j)$ represent samples collected from the posterior of the
chosen $\bm{\phi}$-space (shrinkage) regression models---one of the
ones from Section \ref{sec:bshrink}. 
Then, following
Eq.~(\ref{eq:addy}) from Section \ref{sec:monotone}, samples from the
posterior distribution of $\bm{\theta}$ may be obtained by repeating
the following steps.
\begin{enumerate}
\item Sample $(\mu_1, \Sigma_{1,1}) \equiv (\beta_{0,1}, \sigma_1^2)
  \sim \mathcal{BR}_1$.  See below for details of this special case.
\item For $j=2,\dots,m$:
\begin{enumerate}
\item Sample $(\beta_{0,j}, \bm{\beta}_j, \sigma_j^2) \sim
  \mathcal{BR}_j$.
\item Convert $(\mu_j, \bm{\Sigma}_{1:j,j}) = \Phi^{-1}(\beta_{0,j},
  \bm{\beta}_j, \sigma_j^2, \bm{\mu}_{1:(j-1)},
  \bm{\Sigma}_{1:(j-1),1:(j-1)})$, following Eq.~(\ref{eq:addy}).
\end{enumerate}
\end{enumerate}
Since the Bayesian regressions ($\mathcal{BR}_j$) are mutually
independent, step 1 and the $j-1$ steps of 2(a) may be performed in
parallel, and the conversion via $\Phi^{-1}$ in 2(b) may be performed
offline.  Zeros in $\bm{\beta}_j$ may translate into zeros in
$\bm{\Sigma}_{1:j,j}$ which may be used to test hypotheses about the
marginal and conditional independence between assets.  

In the default formulation of $\mathcal{BR}_j$ with standard normal
errors (\ref{eq:model}) we use $\mb{Y}_j\equiv
\mb{Y}_{1:(j-1)}^{(n_j)}$.  In this case the first step above ($j=1$)
simplifies to:
\begin{align}
\Sigma_{1,1} &\sim \mathrm{IG}\left(\frac{a_\sigma + n_1 - 1}{2}, 
  \frac{b_\sigma +
     ||\tilde{\mb{y}}_1||^2}{2} \right), &&\mbox{then}&
\mu_1 &\sim  \mathcal{N}(\bar{y}_1, \Sigma_{1,1}/n).
\label{eq:nu1prior}
\end{align}
If heavy-tailed errors are modeled [Section \ref{sec:scalemix}] for
the $\bm{\phi}$-space regressions ($\mathcal{BR}_j$), then take
$\mb{Y}_j\equiv \mb{Y}_{0:(j-1)}^{(n_j)}$.  In this case the first
step above ($j=1$) requires integrating over
$\{\omega_{i1}^2\}_{i=1}^{n_1}$.  Conditional on a particular
$\mb{D}_{\omega_1} =
\mathrm{diag}(\omega_{11}^2,\dots,\omega_{n_11}^2)$ sampled from their
full conditional under $\mathcal{BR}_1$ (conditional on $\nu_1$ in
Eq.~(\ref{eq:tpostcond}) which must also be integrated out), we may
sample
\begin{align}
\Sigma_{1,1} &\sim \mathrm{IG}\left(\frac{a_\sigma + n - 1}{2}, 
  \frac{b_\sigma + \mb{y}_1^\top \mb{D}_{\omega_1}^{-1} \mb{y}_1}{2} \right), 
 &&\mbox{then}&
\mu_1 &\sim  \mathcal{N}\left(\frac{\mb{D}_{\omega_1}^{-1} \mb{y}_1}{n_{\omega_1}}, 
  \frac{\Sigma_{1,1}}{n_{\omega_1}}\right),
\label{eq:nu1cond}
\end{align}
where $n_{\omega_1} = {\sum_{i=1}^{n_1} \omega_{i1}^{-2}}$.  Carrying
this through in the second step, above, for $j=2,\dots,m$, etc.,
integrating over independent $(\nu_2, \{\omega_{i2}^2\}_{i=1}^{n_2}),
\dots (\nu_m,\{\omega_{im}^2\}_{i=1}^{n_m})$, is similarly
parallelizable.  The marginal Student-$t$ error structure for
$\mb{y}_j$ carries over into $\bm{\theta}_j$-space giving a distinct
degrees of freedom parameter $\nu_j$ for each (marginal) $\mb{y}_j$,
for $j=1,\dots,m$.  So the resulting model in $\bm{\theta}$-space is
not the typical multivariate Student-$t$ which has a single $\nu$. 

This more standard multivariate Student-$t$ model may be obtained by
modifying the prior so that all $\omega_{ij}^2$ depend on a common
$\nu$.  I.e.,
\[
\omega_{ij}^2 | \bm{\beta}_j, \sigma_j^2, \nu 
\stackrel{\mathrm{iid}}{\sim} \mathrm{IG} 
\left( \frac{\nu + 1}{2}, \frac{\nu + \sigma_j^{-2} 
((\mb{y}_j-\mb{Y}_j \bm{\beta}_j)_i)^2}{2} \right),
\;\;\;\;\; i=1,\dots,n_j, \;\;\;\; j=1,\dots,m 
\]
(using intercept-extended $\mb{Y}_j$ and $\mb{\beta}_j$).  Then, 
the full conditional of $\nu$ becomes
\[
p(\nu|{\{\{\omega_{ij}^2\}_{i=1}^{n_j}\}_{j=1}^m},\theta) 
\propto \left( \frac{\nu}{2} \right) ^{\frac{\nu\sum_{j=1}^m n_j}{2}}
\left(\Gamma \left( \frac{\nu}{2} \right)\right) ^{-\sum_{j=1}^m n_j} 
\exp({-\eta\nu}),
\]
where $\eta = \frac{1}{2}\sum_{j=1}^m\sum_{i=1}^{n_j}
(\log(\omega_{ij}^2)+\omega_{ij}^{-2})+\theta$.  The same rejection
sampling method (i.e., using $n=\sum_{j=1}^m n_j$) applies.  Although
we may proceed with sampling from $\mathcal{BR}_j$ using $\nu_j^{(t)}
\equiv \nu^{(t)}$ ignoring the full conditional for $\nu_j$, the
independence between the $\mathcal{BR}_j$ is now broken:
parallelization of the MCMC is bottlenecked by sampling the common
$\nu$.

The remainder of the section covers the extensions particular to the
portfolio balancing problem: handling ``gaps'', incorporating known
factors, and accounting for estimation risk.

\subsection{Dealing with ``gaps'' by monotone data augmentation}
\label{sec:da}

{\em Data augmentation} (DA) is an established (Bayesian) technique
for dealing with missing data.  In short, it involves treating the
unknown portion of the data as latent variables and updating, or {\em
  imputing}, their values jointly with the other unknown (model)
parameters via the posterior predictive.  For a high level overview
see \citet[][Section 3.4.2]{schafer:1997}.  Rather than treat {\em
  all} of the missing data as latent in this way, it is sufficient
impute a small portion to achieve a monotone missingness pattern.
Then, inference may proceed as already described.  This is known as
{\em monotone data augmentation} (MDA) \citep[][Section
6.5.4]{li:1988,schafer:1997}.  In the case of the financial returns
data $\mb{Y}$ at hand, with sorted columns and rows, the appropriate
candidates for imputation are easily spotted.

Consider each $y_{i,j} = \NA$ such that there exists a $y_{i, j'} \ne
\NA$ where $j' > j$.  Specially mark these with $y_{i,j} = \mathrm{\tt
  NaN}$, say, as these are the entries that must be treated as latent.
There will not be any if the pattern is monotone.  Then sort the rows
of $\mb{Y}$ by the number entries which are (still) {\tt NA} so that
those with more {\tt NA}s appear towards the bottom of $\mb{Y}$.
Define $n_j = \sum_{i=1}^n \mathbb{I}_{\{y_{i,j} \ne \,\mbox{\tt
    \footnotesize NA}\}}$ as before, but now its interpretation is as
the number of observed entries in the \ith{j} column, plus the number
which are treated as latent.  Some entries of $\mb{Y}_j$ and
$\mb{y}_j$, defined as before, may contain {\tt NaN}s.  However,
note that $y_{n_j,j} \ne \mathrm{\tt NaN}$ by construction.

Let $\mb{r}_j$ index the rows of column $j$ of $\mb{Y}$ such that
$\mb{y}_j[\mb{r}_j] = \mbox{\tt NaN}$.  When sampling from
$\mathcal{BR}_j$ in step 2(a), above, ignore the rows of $\mb{Y}_j$
and $\mb{y}_j$ in $\mb{r}_j$, but otherwise proceed as usual.  Then,
add a step, 2(c), wherein for $i \in \mb{r}_j$
\begin{align*}
2(c): && y_{i,j} &\sim \mathcal{N}(\bm{\beta}_j \mb{Y}_{i,1:j}, \sigma_j^2),
\end{align*}
where $\mb{Y}_{i,1:j}$ is the row vector containing the \ith{i} row of
$\mb{Y}_j$.  Observe that the mutual independence of the
$\mathcal{BR}_j$ is broken by this MDA.  They must be processed in
serial so that $\mathcal{BR}_{j+1}(\mb{Y}_{j+1}, \mb{y}_{j+1})$ may
proceed with an up to date copy of $\mb{Y}$.

\subsection{Incorporating known factors}
\label{sec:factor}

A popular way of developing an estimator of the covariance matrix of
financial asset returns is via {\em factor models}.  The idea is that
certain market-level indices, like the value-weighted market index,
the size of the firm associated with the asset, and the book-to-market
factor \cite[e.g.,][]{ckl:1999,famafrench:1993} provide a good basis
for describing individual returns.  Importantly, these factors are
easy to calculate as a function of readily available ``fundamentals''
(characteristics of the listed assets and companies) and the stock
returns.  Through covariances calculated between the factors and
individual asset returns we may infer covariances between each of the
assets.  For a $n \times K$ matrix of (known) factors $\mb{F}$, where
$K \ll \min\{n,p\}$ and where the factors have covariance
$\bm{\Omega}$, the factor model approach poses the following model for
the returns $\mb{Y}$ via columns $j=1,\dots,m$:
\begin{align}
\mb{y}_j &= \lambda_{0,j} + \mb{F}_j \bm{\lambda}_j + \bm{\epsilon}_j, &&
\mbox{where} & \bm{\epsilon}_j &\sim 
\mathcal{N}_{n_j}(\mb{0}, \sigma_j^2\mb{I}_{n_j}).
\label{eq:factor}
\end{align}
Take $\mb{F}_j \equiv\mb{F}_{1:(j-1)}^{(n_j)}$, i.e., without a column
of ones, and treat the regression coefficients $\lambda_{0,j},
\bm{\lambda}_j$ as unknown.  If $\bm{\Lambda}$ is the $K\times m$
matrix defined by collecting the $\bm{\lambda}_j$ column-wise, then a
covariance matrix on the returns $\mb{Y}$ may be obtained as
\begin{align*}
  \bm{\Sigma}^{(f)} &= \bm{\Lambda}^\top \bm{\Omega} \bm{\Lambda} +
  \mb{D}_\sigma, & & \mbox{where} & \mb{D}_{\sigma} &=
  \mathrm{diag}(\sigma_1^2, \dots, \sigma_m^2).
\end{align*}
The MLE $\hat{\bm{\Sigma}}^{(f)}$ may be obtained via the standard
estimate of $\hat{\bm{\Omega}}$ and the MLEs
$\{\hat{\bm{\lambda}}_j\}_{j=1}^m$.  Similarly, one may sample from
the Bayesian posterior with suitable (non-informative) independent
priors on $\mb{\Omega}$, and $\{\lambda_{0,j}, \bm{\lambda}_j,
\sigma_j^2\}_{j=1}^m$.

The estimators of $\mb{\Sigma}^{(f)}$ obtained in this way tend to
have low variance, which is a desirable property.  However, they also
have a strong bias, which may be undesirable.  The bias stems from an
implicit assumption that the returns are mutually independent when
conditioned on the factors.  Not only might this not be a reasonable
assumption, but it also makes the quality of the resulting
estimator(s) extremely sensitive to the choice of factors.  This bias
may be mitigated to some extent by further involving the returns in
the estimation process, i.e., in a more direct way.  One such
approach, considered by \cite{ledoit:2002}, is to take a convex
combination of a factor-based estimator ($\hat{\bm{\Sigma}}^{(f)}$)
and a standard (possibly non-positive definite) complete data
estimator ($\hat{\bm{\Sigma}}^{(c)}$):
\begin{equation}
  \hat{\bm{\Sigma}}^{(\ell)} = \alpha \hat{\bm{\Sigma}}^{(f)} + 
  (1 - \alpha)\hat{\bm{\Sigma}}^{(c)}, 
\;\;\;\;\; \mbox{for } \alpha \in [0, 1].
  \label{eq:ledoit}
\end{equation}
The mixing proportion, $\alpha$, may be determined by CV.  That this
approach works well is a testament to the importance of combining a
factor model with a more direct approach.

\cite{gra:lee:silva:2008} described hybrid method of incorporating the
factors into the $\bm{\phi}$-space procedure of the monotone
factorization MLE so the data may inform on which independence
assumptions are adequate.  Consider the {\em combined} regression
model:
\begin{equation}
\label{eq:combined}
\mb{y}_j = \beta_{0,j} + \mb{Y}_j \bm{\beta}_j + \mb{F}_j
\bm{\lambda}_j + \mb{\epsilon}_j.
\end{equation}
Observe that the $\lambda_{0,j}$ term present in Eq.~(\ref{eq:factor})
has been dropped because it is not identifiable in the presence of
$\beta_{0,j}$.  With some bookkeeping, the model described in
Eq.~(\ref{eq:combined}) can be used to obtain a joint estimator for a
$m+K$ element mean vector and $(m+K) \times (m+K)$ covariance matrix
from which $\bm{\mu}$ and $\bm{\Sigma}$ may be extracted.  To fix
ideas, suppose that the factors are completely observed, which is
usually the case.  Then we may sample from the regression models in
$\mathcal{BR}([\mb{F}_j \; \mb{Y}_j], \mb{y}_j)$ in
$\bm{\phi}$-space, and after transformation to $\bm{\theta}$-space
via $\Phi^{-1}$ in Eq.~(\ref{eq:addy}) the latter $m$ components
$\bm{\mu}_{(K+1):(m+K)}$, and $m$ rows/cols
$\mb{\Sigma}_{(K+1):(m+K),(K+1):(m+K)}$, may be extracted as a sample
of the mean and covariance of the returns.  Shrinkage (or model
averaging) in the regression model enables the columns in $[\mb{F}_j
\; \mb{Y}_j]$, be they factors or returns, that are least useful for
predicting $\mb{y}_j$ to be down-weighted.  That way, rather than
having one parameter governing the trade-off, like $\alpha$ in
Eq.~(\ref{eq:ledoit}), the $m-1$ Bayesian shrinkage regressions can
choose the right balance of factors and returns for each asset.  Prior
knowledge that many assets will be independent when conditioned upon
appropriate factors may reasonably translate into small $\pi$
(controlling the level of sparsity) encoding a preference for a small
proportion of nonzero components of $\bm{\beta}$ in the
$\bm{\phi}$-space regressions.

\subsection{Balancing portfolios and accounting for estimation risk}
\label{sec:pbal}

A portfolio is balanced by choosing $m$ weights $\mb{w}$ describing
the portion of the portfolio invested in each asset.  A standard
technique uses the mean and the covariance between returns to
obtain a {\em mean--variance efficient portfolio}
\citep{markowitz:1959}
by solving a quadratic program (QP).  Common formulations include the
following.\footnote{In all cases we have the tacit constraint that $0
  \leq w_j \leq 1$, for all $j=1,\dots, m$.}

A so-called {\em minimum variance} portfolio may be obtained by
solving
\begin{align}
  \argmin_{\mb{w}} \mb{w}^{\top} \mb{\Sigma} \mb{w}, &&& \mbox{subject
    to} & \mb{w}^\top \mb{1} &= 1.
\label{eq:qpv}
\end{align}
Typical extensions include capping the weights, e.g.,
$0 \leq w_j \leq 2/m$, for $j=1,\dots,m$.

The above formulation may be augmented to involve the estimated mean
return.  One way is to aim for a minimum expected return $\mu$ while
minimizing the variance of the portfolio:
\begin{align}
  \argmin_{\mb{w}} \mb{w}^{\top} \mb{\Sigma} \mb{w} \label{eq:qpmv1},
 && \mbox{subject to}  && \mb{w}^\top \bm{\mu} &\geq \mu,\\
&& \mbox{and} && \mb{w}^\top \mb{1} &= 1. \nonumber
\end{align}
Similar heuristic augmentations apply here as well.  A common
extension is to assume that there is a risk-free asset available,
e.g., a Treasury bond, at rate of return $R_f$.  Then the constraints
may be relaxed to $\bm{\mu} \geq \mu + (1 - \mb{w}^\top \mb{1})R_f
\geq \mu$ and $\mb{w}^\top \mb{1} \leq 1$.  

Given $\bm{\mu}$ and $\bm{\Sigma}$ the solutions to these QPs, which
are strictly convex, are essentially trivial to obtain.
\cite{gra:lee:silva:2008} showed that when shrinkage-based MLEs
$\hat{\bm{\mu}}$ and $\hat{\bm{\Sigma}}$---constructed from all
available returns via the monotone factorized likelihood---are used to
balance portfolios in this way, they outperform a wealth of
alternatives based upon estimators that could only use the completely
observed instances.  In the Bayesian context there are several ready
extensions to this approach.  For example, the MAP parameterization
can be used to balance the portfolio.  Another sensible option is to
use the posterior mean of $\bm{\mu}$ and $\bm{\Sigma}$, which we show
empirically leads to improved estimators of a true (known) generating
distribution [Section \ref{sec:results}], and to improved portfolios
[Section \ref{sec:app}].  But the portfolio balancing problem is about
choosing at time $t$, say, weights to maximize expected return and/or
minimize variance under the posterior predictive distribution
$p(\mb{y}^{(t+1)}| \mb{Y}^{(t)})$.  Here $\mb{Y}^{(t)} \equiv
\mb{Y}_{1:n,:}$ represents the returns available up to time $t$, and
$\mb{y}^{({t+1})}$ is the vector of returns at time
$t+1$.\footnote{The i.i.d.~assumptions erode the meaning of time.
  Notionally, $t$ runs counter to $i=1,\dots,n$.}  {\em Parameter
  uncertainty} (a.k.a., estimation risk) is taken into account by
integration \citep{zellner:chetty:1965,klein:bawa:1976}:
\begin{equation}
p(\mb{y}^{(t+1)}| \mb{Y}^{(t)}) = \int p(\mb{y}^{(t+1)}|\bm{\mu},
\bm{\Sigma}) p(\bm{\mu}, \bm{\Sigma}| \mb{Y}^{(t)}) \, d\bm{\mu}
d\bm{\Sigma}. \label{eq:pred}
\end{equation}
Calculating this integral (or its moments, for use in the QP) is not,
in general, tractable.
However, with i.i.d.~MVN returns (or another elliptical distribution)
the moments are easily obtained \citep{polson:tew:2000} as
$\bm{\mu}^{(t+1)} = \mathbb{E}\{\mb{y}^{(t+1)}|\mb{Y}^{(t)}\} =
\mathbb{E}\{\bm{\mu}|\mb{Y}^{(t)}\}$ and $\bm{\Sigma}^{(t+1)} =
\mathbb{E}\{\bm{\Sigma}|\mb{Y}^{(t)}\} + \mathrm{Var}\{\bm{\mu} |
\mb{Y}^{(t)}\}$, i.e., via a conditional variance identity.

In the case of completely observed returns, a standard non-informative
prior $p(\bm{\mu}, \bm{\Sigma}) \propto
|\bm{\Sigma}|^{-\left(\frac{m+1}{2} \right)}$, and (importantly) more
returns than assets $(n > m)$, \cite{polson:tew:2000} show that there
is ``{\em no effect} of parameter uncertainty on the portfolio rule''
since $\bm{\mu}^{(t+1)} = \hat{\bm{\mu}}$ and $\bm{\Sigma}^{(t+1)} = c
\hat{\bm{\Sigma}}$, where the constant $c$ is available in closed form
and is a function of $n$ and $m$ only.  In the case of historical
returns of varying length, and when $n \gg m$, \cite{stambaugh:1997}
shows that we again have that $\bm{\mu}^{(t+1)} = \hat{\bm{\mu}}$, and
that $\bm{\Sigma}^{(t+1)}$ is available in closed form but is not a
scalar multiple of $\hat{\bm{\Sigma}}$.  It can be shown empirically
that incorporating this parameter uncertainty (a.k.a., estimation risk)
leads to improved investments.

We are motivated by the situations in which these analytical
approaches do not apply, i.e., when $m \geq n$ or when $n_j \leq j$
for any $j=1,\dots,m$.  Although \cite{gra:lee:silva:2008} extended
the MLE approach to the $n_j \leq j$ setting by employing parsimonious
regressions, accounting for estimation risk remained illusive.  The
problem is best exposed in the calculation of Stambaugh's $\tilde{V}$
in Eq.~(69--71), pp.~302, where the resulting diagonal is negative
when $m > n$.  But under the fully Bayesian approach, where samples of
$\bm{\mu}$ and $\bm{\Sigma}$ may be taken from the posterior, the
above conditional variance identity may be used to approximate
$\bm{\Sigma}^{(t+1)}$ with arbitrary precision.

\subsection{Empirical results and comparisons}
\label{sec:results}

As a first point of comparison we pit the MLE based point-estimates
of $\bm{\mu}$ and $\bm{\Sigma}$ against the Bayesian alternative:
posterior expectations.  We simulated synthetic data from known
$\bm{\mu}$ and $\bm{\Sigma}$, imposed a uniformly random monotone
missingness pattern, and then calculated the expected (predictive) log
likelihood (ELL) of the so-parameterized MVN distribution(s).  The
ELL of data sampled from a density $p$ relative to a density $q$
(usually estimated) is given by $ \mathbb{E}_p\{\log q\} = \int p(x)
\log q(x) \;dx = H(p) - D_{\mbox{\tiny KL}}(q
\parallel p)$, where $H(p)= \int p\log p$ is the entropy of $p$, and
$D_{\mbox{\tiny KL}}(q
\parallel p)$ is the Kullback--Leibler (KL) divergence between $q$ and
$p$.  The entropy and KL divergence are known in closed form for MVN
densities $p$ and $q$.  When $q$ uses point-estimates
$(\hat{\bm{\mu}}, \hat{\bm{\Sigma}})$, and $p$ uses the truth
$(\bm{\mu}, \bm{\Sigma})$, the ELL is given by:
\begin{equation}
  -\frac{1}{2} \log \{(2\pi e)^N |\bm{\Sigma}|\} 
  -\frac{1}{2} \left(\log
  \frac{|\hat{\bm{\Sigma}}|}{|\bm{\Sigma}|} +
  \mbox{tr}(\hat{\bm{\Sigma}}^{-1} \bm{\Sigma}) + (\hat{\bm{\mu}} -
  \bm{\mu})^\top \hat{\bm{\Sigma}}^{-1}(\hat{\bm{\mu}} - \bm{\mu}) \right).
\label{eq:ell}
\end{equation}

\begin{table}[ht!]
\centering
\footnotesize
\begin{tabular}{r||rrr|rrr|rrr||rrr|rrr}
  & \multicolumn{9}{c||}{Fully Bayesian} & \multicolumn{6}{c}{MLE/CV}
  \\
  & \multicolumn{3}{c|}{NG}
  & \multicolumn{3}{c|}{Lasso} & \multicolumn{3}{c||}{Ridge} &
  \multicolumn{3}{c|}{Lasso} & \multicolumn{3}{c}{Ridge} \\
  $\delta$ & 0.9
  & 0.2 & 0 & 0.9 & 0.2 & 0 & 0.9 & 0.2 & 0 & 0.9 & 0.2 & 0 & 0.9 & 0.2 & 0 \\
  \hline \hline
  & \multicolumn{15}{c}{{\tt normwish}, $m=100$, $n=100$} \\
  \hline
  min & 8 & 3 & 3 & 8 & 3 & 3 & 3 & {\gr 1} & 1 & 10 & 10 & 10 & {\rd 12} & 11 & 10 \\
  mean & 8.5 & 4.4 & 5.1 & 8.5 & 3.7 & 4.9 & 6.8 & {\gr 1.4} & 1.6
  & 13.1 & 11.6 & 11.7 & {\rd 14.0} & 12.7 & 11.9  \\
  max & 9 & 7 & 7 & 9 & 6 & 7 & 7 & {\gr 2} & 2 & 15 & 15 & 15 & {\rd 15} & 14 & 15 \\
  \hline
  & \multicolumn{15}{c}{{\tt parsimonious}, $m=100$, $n=100$} \\
 \hline
 min & 7 & 2 & {\gr 1} & 7 & 3 & 1 & 9 & 5 & 5 & 10 & 7 & 6 & {\rd 13} & 13 & 12 \\
 mean & 7.7 & 3.4 & {\gr 1.3} & 8.1 & 3.6 & 1.7 & 9.6 & 5.9 & 5.1 & 12.0 &
 10.2 & 9.5 & {\rd 14.9} & 13.8 & 13.1 \\
 max & 10 & 4 & {\gr 2} & 10 & 4 & 3 & 11 & 7 & 6 & 15 & 15 & 12 & {\rd 15} & 14 & 15 \\
  \hline
  & \multicolumn{15}{c}{{\tt normwish}, $m=100$, $n=1000$} \\
  \hline
  min & 7 & 1 & 3 & 7 & 1 & 3 & 5 & {\gr 1} & 1 & 10 & 10 & 10 & {\rd 12} & 11 & 10 \\
  mean & 8.4 & 3.4 & 5.4 & 8.5 & 3.4 & 5.5 & 7.0 & {\gr 1.2} & 2.2 & 13.7 & 11.5 & 11.7 & {\rd 14.2} & 12.5 & 11.5 \\
  max & 9 & 6 & 6 & 9 & 6 & 7 & 9 & {\gr 4} & 4 & 15 & 15 & 15 & {\rd 15} & 14 & 13 \\
 \hline
  & \multicolumn{15}{c}{{\tt parsimonious}, $m=100$, $n=1000$} \\
  \hline
  min & 7 & 4 & {\gr 1} & 7 & 3 & 1 & 7 & 6 & 3 & 11 & 7 & 4 & {\rd 12} & 11 & 10 \\
  mean & 8.8 & 4.3 & {\gr 1.1} & 9.1 & 4.8 & 1.9 & 10.2 &  6.1 &  3.0 & 12.1 & 9.2 & 7.9 & {\rd 14.9} & 13.8 & 12.8 \\
  max & 11 & 6 & {\gr 2} & 11 & 5 & 2 & 11 & 7 & 4 & 15 & 15 & 15 & {\rd 15} & 14 & 13
\end{tabular}
\vspace{0.2cm}
\caption{Summary of the rankings (by ELL (\ref{eq:ell})) of fifteen 
  parsimonious regression methods
  used in the monotone MVN algorithm on 50 randomly generated MVN 
  parameterizations, via two data generation ``{\tt method}s''.
  The best ranks appear in green; worst in red. }
\label{t:ell}
\end{table}

Table \ref{t:ell} contains summary information about the relative
rankings of the ELL calculations for nine Bayesian and nine MLE
estimators in a series of 50 repeated experiments.  The results in the
top portion of the table are for random MVN parameterizations that
were obtained using the {\tt randmvn} function from the {\tt monomvn}
package, with argument \verb!method="normwish"!, and $m=n=100$.
Uniform monotone missingness patterns are then obtained with {\tt
  rmono}. The use of parsimonious $\bm{\phi}$-space regressions
(lasso, NG, ridge) with model selection via RJ-MCMC is determined by
the setting of $\delta\in[0,1)$.  A parsimonious regression is
performed if $\delta n_j < j$, and OLS is used otherwise---OLS
regression is never used when $\delta=0$.  Observe from this portion
of the table that the Bayesian estimators always obtain a better rank
than the MLE ones.  In the Bayesian case, ridge regression nudges out
the lasso/NG.  This makes sense because the data generation method
(\verb!"normwish"!) does not allow for any marginal or conditional
independencies, i.e., zeros in $\bm{\Sigma}$ or $\bm{\Sigma}^{-1}$.
The situation is reversed for the MLE, where the lasso comes out on
top.  
In all cases but the MLE/CV lasso implementation, a lower value of
$\delta$ gives improved performance, indicating that parsimonious
regressions yield improvements even when they are not strictly
necessary.  However, lower $\delta$ comes with a higher computational
cost. Given that the improvement of $\delta=0$ over $\delta=0.2$ is
slight, the higher setting may be preferred.  Finally, observe that
the Bayesian lasso edges out the NG.

The results of a similar experiment, except using the argument
\verb!method="parsimonious"!  to {\tt randmvn}, are summarized in the
second portion of the table.  This data generation mechanism allows
for conditional and marginal independence in the data by building up
$\bm{\mu}$ and $\bm{\Sigma}$ sequentially, via randomly generating
$\bm{\beta}_j$ (possibly having zero-entries) and applying $\Phi^{-1}$
as in Eq.~(\ref{eq:addy}).  Here the number of nonzero entries of
(each) $\bm{\beta}_j$ follows a Bin$(j, 0.1)$ distribution.  In this
case the results in the table indicate that the lasso/NG is the
winner. The distinction between Bayesian and MLE/CV methods is as
before.  This time, NG edges out the Bayesian lasso. Finally, the
bottom two portions of the table report ranks for the same two
experiments described above, but using $n=1000$ so that parsimonious
regressions are needed less often. These results are similar to the
$n=100$ case except that the distinction between the Bayesian and
MLE/CV ranks is blurred somewhat.

\section{Asset management by portfolio balancing}
\label{sec:app}

Here we return to the motivating asset management problem from Section
\ref{sec:intro}.  We examine the characteristics of minimum variance
portfolios (\ref{eq:qpmv1}) constructed using estimates of
$\bm{\Sigma}$ based upon historical monthly returns through a Monte
Carlo experiment of repeated investment exercises.  
The experimental setup closely mirrors the one used by
\cite{gra:lee:silva:2008}, modeled after \cite{ckl:1999}.  The data
consist of returns of common domestic stocks traded on the NYSE and
the AMEX from April 1968 until 1998 that have a share price greater
than \$5 and a market capitalization greater than 20\% based on the
size distribution of NYSE firms.  All such ``qualifying stocks'' are
used---not just ones that survived to 1998.  Since the
i.i.d.~assumption is only valid locally (in time) due to the
conditional heteroskedastic nature of financial returns, estimators of
$\bm{\Sigma}$ are constructed based upon (at most) the most recently
available 60 months of historical returns.  Short selling is not
allowed; all portfolio weights must be non-negative.  Although
practitioners often impose a heuristic cap on the weights of balanced
portfolios, e.g., at 2\%, in order to ``tame occasional bold
forecasts'' \citep{ckl:1999} or to ``curb the effects \dots\ of poor
estimators'' \citep{jagma:2003}, we specifically do not do so here in
order to fully expose the relative qualities of the estimators in
question.

Our analysis in Section \ref{sec:scalemix} suggests that the benefit
of modeling Student-$t$ errors will lead to minor improvements in our
estimators based upon just 60 or fewer historical returns.
\cite{jacq:polson:rossi:2004} showed that Bayes factors give strong
preference to (the simpler) normal model over the Student-$t$ at
return frequencies less than weekly---supporting the {\em aggregation
  normality} of monthly returns.  Models with Student-$t$ errors were
included in initial versions of our exercise, but they performed no
better than their normal counterparts.  This, and the extra
computational burden required by the extra $O(n)$ extra latent
variables, led us to exclude the Student-$t$ comparators from the
experiment reported on below.

The Monte Carlo experiment consists of 50 random repeated paths
through 26 years, starting in April 1972.  In each year 250
``qualifying stocks'' with at least 12 months of historical returns
are chosen randomly without replacement.  Using at most the last 60
returns of the 250 assets, estimates of the covariance matrix
$\bm{\Sigma}$ of monthly excess returns (over the monthly Treasury
Bill rate) are calculated under our various methods and used to
construct minimum variance portfolios.  The portfolios are then held
(fixed) for the year.  To assess their quality and characteristics we
follow \cite{ckl:1999} in using the following: (annualized) mean
return and standard deviation; (annualized) Sharpe ratio (average
return in excess of the Treasury bill rate divided by the standard
deviation); (annualized) tracking error (standard deviation of the
return in excess of the S\&P500); correlation to the market (S\&P500);
average number of stocks with weights above 0.5\%.  Generally
speaking, portfolios with high mean return and low standard
deviation, i.e., with large Sharpe ratio, are preferred.  Sharpe
ratios being roughly equal, we prefer those with lower tracking
error. 

\begin{table}[ht!]
\vspace{0.2cm}
\begin{center}
\begin{tabular}{r||rrrrrr}
 & mean & sd & sharpe & te & cm & wmin \\
  \hline \hline
eq & 0.149 & 0.188 & 0.431 & 0.063 & 0.949 & 0 \\
  vw & 0.134 & 0.162 & 0.404 & 0.032 & 0.980 & 45 \\
\hline
  com & 0.151 & 0.182 & 0.457 & 0.107 & 0.812 & 26 \\
  min & 0.150 & 0.183 & 0.448 & 0.106 & 0.816 & 29 \\
  rm & 0.131 & 0.130 & 0.486 & 0.095 & 0.802 & 16 \\
\hline
  fmin & 0.141 & 0.146 & 0.498 & 0.085 & 0.844 & 39 \\
  fcom & 0.143 & 0.146 & 0.509 & 0.087 & 0.840 & 38 \\
  frm & 0.136 & 0.130 & 0.519 & 0.117 & 0.685 & 21 \\
\hline
  ridge & 0.158 & 0.165 & 0.540 & 0.122 & 0.717 & 18 \\
  bridge & 0.140 & 0.129 & 0.554 & 0.089 & 0.829 & 27 \\
  lasso & 0.149 & 0.149 & 0.543 & 0.054 & 0.940 & 69 \\
  blasso & 0.144 & 0.136 & 0.561 & 0.078 & 0.871 & 39 \\
  bng & 0.144 & 0.136 & 0.560 & 0.078 & 0.872 & 39 \\
\hline
  fridge & 0.158 & 0.164 & 0.549 & 0.121 & 0.719 & 19 \\
  bfridge & 0.142 & 0.129 & 0.571 & 0.085 & 0.846 & 34 \\
  flasso & 0.150 & 0.148 & 0.552 & 0.056 & 0.935 & 69 \\
  bflasso & 0.148 & 0.138 & 0.573 & 0.070 & 0.898 & 51 \\
  bfng & 0.148 & 0.138 & 0.574 & 0.071 & 0.896 & 50
\end{tabular}
\end{center}
\vspace{-0.3cm}
\caption{Comparing statistics summarizing the returns of
  yearly buy-and-hold portfolios generated over 50 repeated 
  random paths through the 26 years of monthly historical returns.
} \label{t:sharpe}
\end{table}

Table \ref{t:sharpe} summarizes the results.  It is broken into five
sections, vertically.  The first section gives results for the equal-
and value-weighted portfolios.  The second section uses standard
estimators of $\bm{\Sigma}$ based only upon the complete data.  The
``min'' estimator uses only the last 12 months of historical returns,
whereas ``com'' uses the maximal complete history available.  Both use
standard estimators (i.e., via the {\tt cov} function in {\sf R}).
The ``rm'' estimator is similar but discards any assets without the
full 60 months of historical returns.  These three rows in the table
highlight that the more historical returns (within the five-year
window) that can be used to estimate covariances the better.  The
third section, containing the same acronyms with a leading ``f'',
incorporates the value-weighted factor on same subset of returns.
The improved characteristics in the table show that good factors can
be quite helpful.

The results are further improved when the estimators exploit the
tractable factorization of the likelihood under the monotone
missingness pattern using shrinkage regression ($\delta = 0.2$), as
shown in the penultimate section of the table.  Notice that the Sharpe
ratios indicate that the fully Bayesian estimators (with a ``b''
prefix) outperform the classical alternative, and that the lasso/NG
methods are better than the ridge.  In addition to fully accounting
for all posterior uncertainties, the Bayesian estimators have the
advantage of being able to deal with ``gaps'' in the data, via MDA
[Section \ref{sec:da}], and can account for estimation risk [Section
\ref{sec:pbal}].  Observe that these estimators distribute the weight
less evenly among the assets, having fewer assets with $\geq 0.5\%$ of
the weight on average compared to their classical counterparts.  The
higher concentration of weight on the appropriate assets leads to
lower variance portfolios which deviate further from the market, hence
the somewhat higher tracking error.  The results for the lasso are
nearly identical to those under the extended NG
formulation. 

The final section of the table shows that incorporating the
value-weighted factor (now with $\delta = 0$) leads to further
improvements.  A sensible (prior) belief that the presence of a good
factor causes many pairs of assets to be conditionally independent
[Section \ref{sec:pbal}] allows us to dial down the hierarchical prior
on the proportion of nonzero regression coefficients: $\pi\sim
\mathrm{Beta}(1, 100)$.  The results in the table suggest that the
factor causes weight to be more evenly distributed in the case of the
Bayesian estimators, but not the classical ones.

The first eight rows (first three sections) of the table, and those
corresponding to ``lasso'', ``ridge'', ``flasso'' and ``fridge'', are
nearly identical to ones in a similar table from
\cite{gra:lee:silva:2008}.  Any variation is due to different random
seeds.  This calibration allows us to draw comparisons to the other
classical {\tt monomvn} estimators including ones based upon PCR, etc.
In short, the fully Bayesian approach(es) reign supreme.  The
improvements may appear to be modest at a glance.  But in light of the
fact that financial markets are highly unpredictable they are actually
quite substantial.  As a matter of curiosity we also calculated
statistics under the posterior mean parameterization, i.e., without
accounting for estimation risk.  The Sharpe ratios were: 0.549 (0.562
with the factor) under the ridge, 0.554 (0.562) under the lasso, and
0.553 (0.563) under the NG.  These numbers point to an improvement
over the classical approach using the posterior mean, but indicate
that the incorporation of estimation risk is crucial to get the best
portfolio weights.

\begin{figure}[ht!]
\centering
\includegraphics[trim=20 0 0 10,scale=0.85]{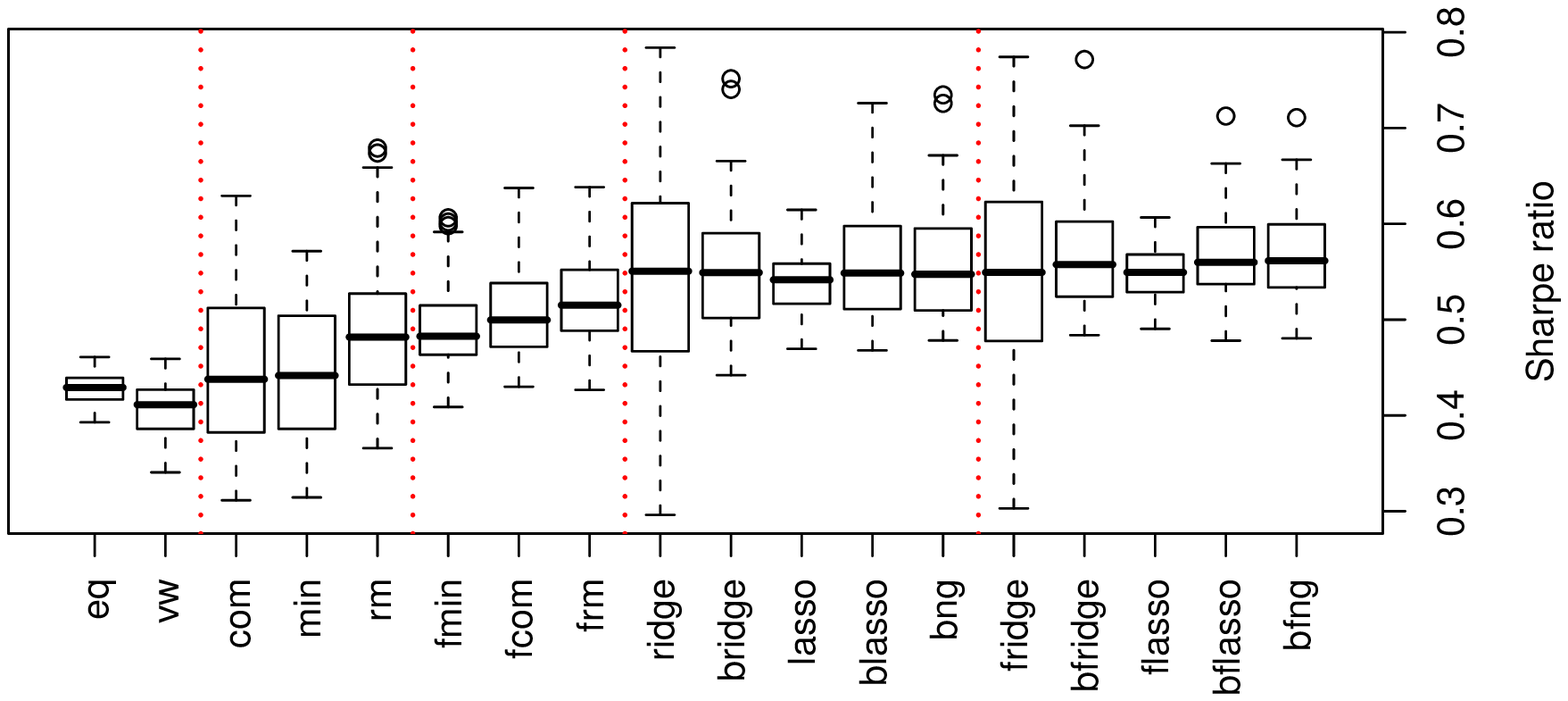}
\includegraphics[trim=20 0 0 25,scale=0.85]{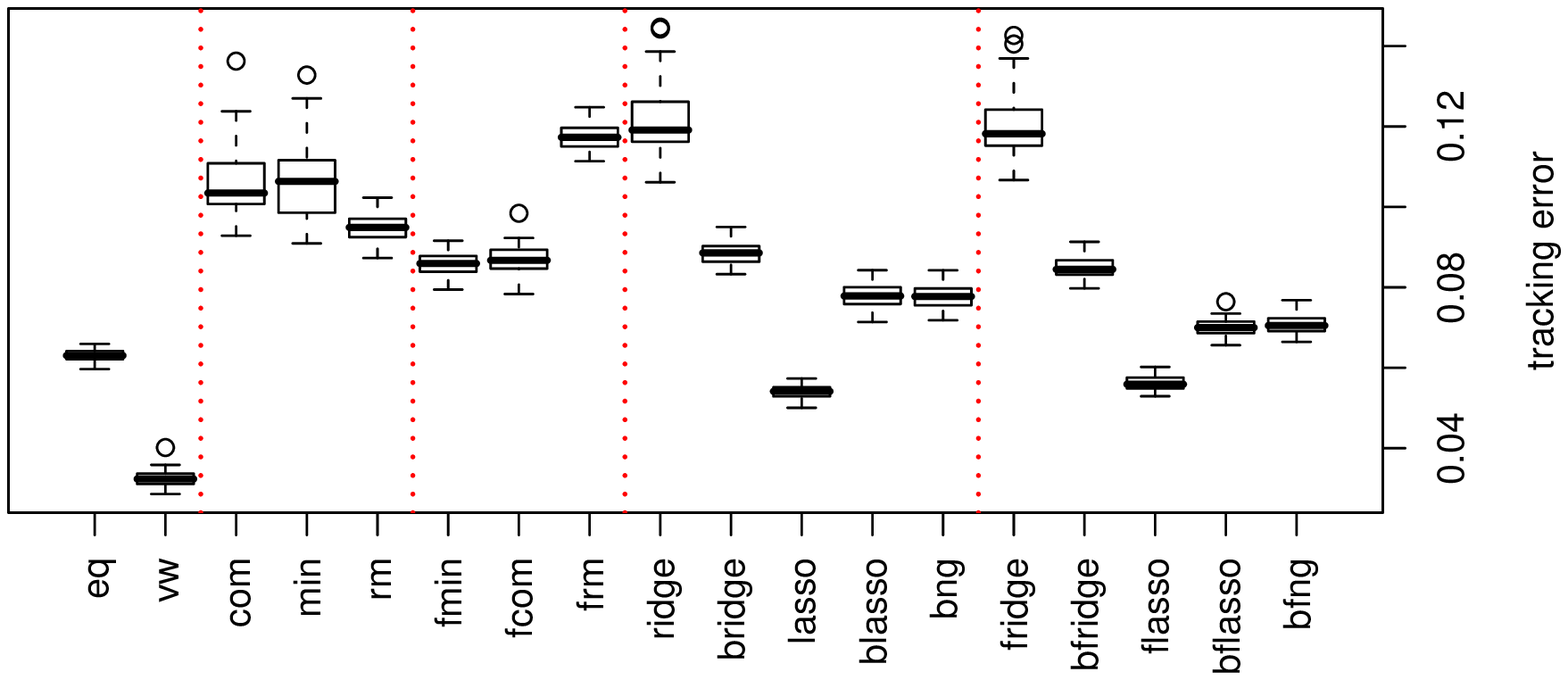}
\caption{Boxplots of Sharpe ratios {\em (top)} and the tracking error
  {\em (bottom)} obtained over 50 random paths through 26 years.
  The vertical bars correspond to horizontal ones in Table
  \ref{t:sharpe}.}
\label{f:boot}
\end{figure}

Figure \ref{f:boot} summarizes the variability in the Monte Carlo
experiment showing the distribution (via boxplots) of the Sharpe
ratios and tracking error obtained for each of the 50 random paths
through the 26 years, thereby complementing the averages presented in
Table \ref{t:sharpe}.  We can immediately see that the classical ridge
regression approach is highly variable, often yielding extremely low
Sharpe ratios and high tracking error.  The Bayesian approach offers a
dramatic improvement here.  In the case of the lasso/NG we can see
that the variability of the Sharpe ratios for the Bayesian
implementations are higher than their classical counterparts.
However, it is crucial to observe that the boxplots extend in the
direction of larger Sharpe ratios, offering improved estimators.

\section{Discussion}
\label{sec:discuss}

We have shown how the classical (MLE/CV) shrinkage approach of
\cite{gra:lee:silva:2008} to joint multivariate inference under
monotone missingness may be treated in a fully Bayesian way to great
effect.  The Bayesian approach facilitates extensions to deal with
``gaps'' in the monotone pattern via MDA and heavy-tailed data, and
can account for estimation risk.  None of these features could be
accommodated by the classical approach.  In synthetic data experiments
we demonstrated the descriptive, predictive, and inferential
superiority of the Bayesian methods.  Using real financial returns
data we showed how the fully Bayesian approach leads to portfolios
with lower variability and thus higher Sharpe ratio.

Our methods bear some similarity to other recent approaches to
covariance estimation.  \citet{lev:roth:zhu:2008} and
\citet{carvalho:scott:2009} offer priors on covariance matrices with
shrinkage based on Cholesky decompositions. Our monotone factorization
\citep[originally:][]{andersen:1957} can be seen as a special case
where the column order, and thus the underlying graphical model
structure, is fixed by historical availability.  This has certain
advantages in our context (relative inferential simplicity,
tractability, MDA and heavy tail extensions), but it may not be
optimal in complete data cases.  \cite{liu:1996,liu:1995} provides a
model for Bayesian robust multivariate joint inference for data
exhibiting a monotone missingness pattern.  The posteriors are derived
using extensions of Bartlett's decomposition, and ``gaps'' may be
handled via MDA.  Although a wealth of theoretical and empirical
results are provided, it is not clear how the methods can be adapted
to ``big $p$ small $n$'' setting.  A possible way forward involves
Bayesian dynamic factor models \citep{west:2003} which are designed
for ``big $p$ small $n$'' and retain the ability to handle missing
data in a tractable way.

It is becoming well understood that the Laplace prior has many
drawbacks, even when generalized by the NG.  For example, it is known
to produce biased estimates of the nonzero coefficients and to
underestimate of the number of zeros.  A newly developed shrinkage
prior for regression called the horseshoe
\citep{carvalho:polson:scott:2008} shows promise as a tractable
alternative (i.e., via GS) without the bias problems. Incorporating
horseshoe regression in our framework is part of our future
work. Another obvious extension is to relax the i.i.d.~assumption to
obtain more dynamic estimators.  One possible approach might be to use
weighted regressions for the $\mathcal{BR}$'s with weights decaying
back in time.

\subsection*{Acknowledgments} This work was partially supported by
Engineering and Physical Sciences Research Council Grant EP/D065704/1.
We would like to thank an anonymous referee, and the associate editor,
whose many helpful comments improved the paper.


\bibliography{corr}
\bibliographystyle{jasa}

\end{document}